\documentclass[twocolumn,aps,pra,superscriptaddress,footinbib,floatfix]{revtex4-2}
\usepackage[utf8]{inputenc}

\usepackage{amsmath,amssymb}
\usepackage{hyperref,xcolor,graphicx}

\usepackage{tikzsymbols}

\renewcommand{\Im}{{\rm Im}\,}
\renewcommand{\Re}{{\rm Re}\,}

\newcommand{\ket}[1]{|#1\rangle}
\newcommand{\bra}[1]{\langle #1|}
\newcommand{\braket}[2]{\langle #1|#2\rangle}
\newcommand{\braOket}[3]{\langle #1|#2|#3\rangle}

\begin{document}
\title{Ideal Chern bands are Landau levels in curved space}
\author{Benoit Estienne}
\affiliation{Sorbonne Universit\'e, CNRS, Laboratoire de Physique Th\'eorique et Hautes \'Energies, LPTHE, F-75005 Paris, France}
\author{Nicolas Regnault}
\affiliation{Laboratoire de Physique de l'Ecole normale sup\'{e}rieure, ENS, Universit\'{e} PSL, CNRS, Sorbonne Universit\'{e}, Universit\'{e} Paris-Diderot, Sorbonne Paris Cit\'{e}, 75005 Paris, France}
\affiliation{Department of Physics, Columbia University, New York, NY 10027, USA}
\author{Valentin Cr\'epel}
\affiliation{Center for Computational Quantum Physics, Flatiron Institute, New York, New York 10010, USA}

\begin{abstract}
We prove that all the criteria proposed in the literature to identify a Chern band hosting exact fractional Chern insulating ground states, in fact, describe an equivalence with a lowest Landau level defined in curved space under a non-uniform magnetic field. In addition, we design an operational test for the most general instance of such lowest Landau level mapping, which only relies on the computationally inexpensive evaluation of Bloch wavefunctions' derivatives. Our work clarifies the common origin of various Chern-idealness criteria, proves that these criteria exhaust all possible lowest Landau levels, and hints at classes of Chern bands that may posses interesting phases beyond Landau level physics.
\end{abstract}

\maketitle

\paragraph*{Introduction --- } The remarkable promise of fractional Chern insulators (FCIs)~\cite{tang2011high,neupert2011fractional,sheng2011fractional,regnault2011fractional}, is to realize the universal anyonic physics existing in fractional quantum Hall (FQH) systems~\cite{stormer1999fractional,crepel2018matrix,hansson2017quantum,crepel2019matrix,jain1990theory,jain2015composite,crepel2019microscopic,moore1991nonabelions,stern2008anyons,crepel2019model,clarke2013exotic,crepel2019variational,arovas1984fractional,simon1998chern,crepel2020microscopic} using less stringent experimental conditions, and in particular, without any external magnetic field. At the same time, FCIs remain fragile strongly correlated phases of matter, and abundant theoretical research efforts aimed at identifying propitious conditions for their emergence. Arguably, some of the most promising insights towards that goal originated from the realization that $|C|=1$ Chern bands with constant quantum geometric tensor (QGT) reproduce the interaction form factors of the standard lowest Landau level (LLL) wavefunctions on the torus~\cite{roy2014band}. As a result, the interacting physics of both systems are identical, allowing to transpose most of our analytical understanding of FQH systems to such bands. This LLL-mapping argument identifies certain bands in which analytical arguments can ensure the emergence of FCIs, and can serve as a guide in our long-standing search for zero fields analogs of FQH states.

Motivated by recent progress in twisted bilayer graphene~\cite{tarnopolsky2019origin,vafek2020renormalization,bultinck2020ground,lian2021twisted,bernevig2021twisted,crepel2023chiral2} and moir\'e semiconductors~\cite{crepel2022anomalous,crepel2023chiral}, more general conditions have been obtained to capture a larger set of bands in which FCIs appear as the ground state of specific interacting Hamiltonian at fractional filling. These various criteria for idealness have introduced a zoology of special Chern bands, those of constant QGT carrying the name ``flat K\"ahler bands''~\cite{claassen2015position,mera2021engineering}, which were joined by ``ideal bands'' featuring a non-uniform QGT that nevertheless possesses a constant null vector~\cite{wang2021exact}, which were themselves extended to so-called ``vortexable bands'' allowing for non-linear embedding in real space~\cite{ledwith2022vortexability}. Each of these classes is defined by the ability to obtain a model interacting Hamiltonian for FCIs, akin to the pseudopotentials in FQH systems~\cite{haldane1983fractional}.

This raises the question whether these apparently different classes of Chern bands have a common physical origin, and whether the number classes will keep growing as new model Hamiltonians are found or will instead converge to a complete definition. In this work, we convey concrete answers about these two questions.

First, we prove that all currently existing criteria for special bands with unit Chern number are generalized version of LLL-mapping, accounting for a non-uniform magnetic field and cyclotron metric. In particular, ideal bands allow for a non-uniform Berry curvature, which is equivalent to the physics of LLLs with non-uniform magnetic field, as already hinted, for instance, in Refs.~\cite{douglas2010bergman,konstantinou2017forgotten,dong2022dirac}. Vortexable bands describe how the density profile of Bloch states changes near their zeros at different points in space, which simply corresponds to LLLs equipped with a non-uniform cyclotron metric. Second, we conversely demonstrate that the most general class of periodic LLL can be reproduced by at least one of the special Chern bands so far introduced, bolstering the generality of Ref.~\cite{ledwith2022vortexability}. An ideal Chern band, in the most liberal sense, is therefore in one-to-one mapping with a LLL, proving that there cannot exist more general criteria describing Chern idealness by analogy with LLLs. We thirdly extend the demonstration to higher Chern numbers using their decomposition into colors~\cite{wu2014haldane}. For practical purposes, we finally design an operational criterion to test which Chern bands can be mapped a LLL with non-uniform magnetic field and metric. Our criterion only involves derivatives of the Bloch eigenfunctions, and therefore remains computationally inexpensive.

\paragraph*{Most general periodic lowest Landau level --- } The Landau problem describes the quantization of cyclotron orbits of free massive charged particles in a magnetic field $B = \varepsilon^{ab} \partial_a A_b$ on a two-dimensional plane. In the simplest scenario where the mass tensor and magnetic field are uniform, the spectrum is composed of Landau levels (LLs), which are equally spaced energy levels with extensive degeneracy due to the unconstrained guiding center degree of freedom. On allowing for a non-uniform mass tensor or a curved plane, a metric $g_{ab}$ is introduced, but the spacing and degeneracy of LLs are not compromised as long as the magnetic flux density $B/\sqrt{g}$ remains constant, where $\sqrt{g} = \sqrt{\det g_{ab}}$~\cite{alicki1993landau,klevtsov2014random,klevtsov2016geometry,klevtsov2017lowest}. In all other cases, LLs generically exhibit a finite dispersion. However, the LLL can still be flattened and locked at zero energy by adding an electric potential $V = B/(2 \sqrt{g})$~\cite{dubrovin1980ground}. Therefore, the most generic flat LLL on a plane can be obtained by considering the ground-state manifold $\mathcal{L}(g,B)$ of the Hamiltonian
\begin{equation} \label{eq_genericLandau}
\mathcal{H}(g,B)= \frac{1}{2 \sqrt{g}} \left[\pi_a g^{ab} \sqrt{g} \pi_b -  B \right] \, , 
\end{equation}
with $\pi_{a} = - i \partial_a - A_a$ the canonical momentum (we use units in which the Planck's constant, the electric charge and electric mass are equal to one $\hbar=e=m=1$).

To capture the physics of periodic two-dimensional systems, we convert these LLL into a band problem. For this to happen, the metric and the magnetic field must be periodic with respect to a Bravais lattice $\mathbb{Z} a_1 + \mathbb{Z} a_2$, with an integer number of flux quanta $N_\phi$ per unit cell. This last requirement ensures that the magnetic translations $T_{j}$ by $a_{j}$ for $j=1,2$ commute and can be simultaneously diagonalized. The number of bands obtained via this construction is equal to $N_\phi$. Since we are interested in describing a single band, we fix $N_\phi=1$ from now on.

To find the LLL of $\mathcal{H}(g,B)$, it is convenient to use isothermal coordinates $(X,Y)$ in which the metric is $ds^2 = e^{2\sigma (X,Y)} (dX^2 + dY^2)$ with $\sigma$ a real function~\cite{gauss1825allgemeine}. The complex coordinate $Z = X + i Y$ can be chosen such that $Z(r+a_1) = Z(r) + 1$ and $Z(r+a_2) = Z(r) + \tau$, where $\tau$ is a complex number, and $r = (x,y)$ denotes the original coordinate system (see App.~\ref{app_GlobalIsothermal}). In isothermal coordinates the Landau Hamiltonian becomes $\mathcal{H}(g,B) = \frac{1}{2 m} e^{-2 \sigma} \Pi \,\overline{\Pi}$ where $\Pi = \pi_X + i \pi_Y$, and LLL wavefunctions are those annhilated by $\bar{\Pi}$. This property is conformally invariant, in the sense that it is not sensitive to the conformal factor $e^{2\sigma}$. Thus LLL wavefunctions are invariant, up to normalization, under Weyl transformation (local rescaling of the metric).

To be more explicit, we split the total magnetic field as $B = B_0 + \tilde{B}$ where $B_0 = 2 \pi / \Im\tau$ is uniform in isothermal coordinates and $\tilde{B}$ carries no flux on the unit cell. The LLL wavefunction $\ket{\psi_k} \in \mathcal{L}(g,B)$ diagonalizing the magnetic translations $T_j \ket{\psi_k} = e^{ik_j} \ket{\psi_k}$ takes the first-quantized form (see App.~\ref{app_FormLLLelements})
\begin{equation} \label{eq_mostgeneralLLL} 
\psi_k (x,y) = e^{-\tilde{\rho}(X,Y)} \phi_k (X,Y) / N_k ,  
\end{equation}
where ($i$) $\ket{\phi_k} \in \mathcal{L}(\delta_{ab}, B_0)$ denotes the wavefunction of the LLL defined on the torus $\mathbb{C}/({\mathbb{Z} + \tau \mathbb{Z}})$ with uniform and isotropic metric, and displaying the same magnetic translation eigenvalues as $\ket{\psi_k}$, ($ii$) $\tilde{\rho}$ stands for the periodic Poisson (or K\"ahler) potential representing the fluctuating part of the magnetic field, \textit{i.e.} $\Delta \tilde{\rho} = \tilde{B} / \sqrt{g}$ in the original metric, and ($iii$) $N_k^2 = \int {\rm d} X {\rm d} Y |e^{\sigma - \tilde{\rho}} \phi_k|^2 $ is the normalization constant carrying information about the local space curvature. For completeness, we also provide the explicit form of flat and uniform LLL functions (see App.~\ref{app_FormLLLelements})
\begin{equation} \label{eq_flatuniformLLL}
\phi_k (X,Y) = e^{\frac{B_0}{4} Z (Z - \bar{Z}) + i (k_1 +\pi) Z } \theta_1 (Z-Z_k; \tau) ,
\end{equation}
whose zero is fully determined by the quasi-momentum that fixes $2 \pi Z_k = (k_2 +\pi) - \tau  (k_1 + \pi)$, where we have used the symmetric gauge and denoted as  $\theta_1$ the first Jacobi theta function.

To conclude, the most generic LLL $\mathcal{L}(g,B)$ can be written as that of the conformally equivalent flat surface using isothermal coordinates, with the addition of a periodic Poisson field and a scalar product that correctly includes the Jacobian of the transformation to the new coordinates.

\paragraph*{Equivalence with ideal bands --- } In this manuscript, our first goal is to show that any ideal Chern band (for all flavors of idealness introduced in the literature) maps to a LLL $\mathcal{L}(g,B)$ for some choice of non-uniform field and metric. In the present context, mapping means that the scalar product and form factors computed within the Chern band can be reproduced using LLL wavefunctions from Eq.~\ref{eq_mostgeneralLLL}, ensuring that both systems share the same interacting phase diagram. This definition allows the single-particle wavefunctions of both systems to solely differ by an overall spatially-dependent phase factor that does not change the values of overlaps, \textit{i.e.} for all $\psi_{\rm C}$ in the ideal Chern band there exists $\psi_{\rm L}$ in $\mathcal{L}(g,B)$ such that $\psi_{\rm C} = e^{is} \psi_{\rm L}$ with $s$ a real function independent of $\psi_{\rm C}$. This phase factor is needed to reconcile the respective periodic and the quasi-periodic magnetic boundary conditions of $\psi_{\rm C}$ and $\psi_{\rm L}$. Equivalence up to this boundary condition sewing phase factor shall be written $\psi_{\rm C} \equiv \psi_{\rm L}$ in the rest of this work.

For the sake of presentation, we split the discussion into two parts. We first focus on the momentum-space conditions for idealness involving the quantum geometric tensor~\cite{roy2014band,wang2021exact}, before turning to the more recent and more general real-space criterion relying on the existence of a ``vortex'' function~\cite{ledwith2022vortexability}. We also focus on the case $C=1$, relegating the discussion of greater Chern number to the end of the paper and App.~\ref{app_mostgenericcolorcentered}.

\underline{Momentum-space condition.} Consider a $C=1$ Chern band spanned by Bloch wavefunctions $\ket{\psi_k}$ of unit-cell periodic parts $\ket{u_k} = e^{- i k\cdot r} \ket{\psi_k}$, the quasi-momentum $k$ fixing the eigenvalues under the two elementary translations of the lattice $\psi_k(r + a_j) = e^{i k_j} \psi_k(r)$. Then, define the quantum geometric tensor as $\mathcal{Q}_{k}^{ab} = \braket{D_k^a u_k}{D_k^b u_k}$ using the covariant derivative $D_k^a = \partial_k^a - \braOket{u_k}{\partial_k^a}{u_k}$. The band is said to be $q$-ideal if $\mathcal{Q}$ possesses a constant null vector $w$ throughout the Brillouin zone: $\mathcal{Q}_k^{ab} w_b = 0$ for all $k$.

The relation between $q$-ideal bands and LLs has already been acknowledged~\cite{wang2021exact,dong2022dirac}, and follows from two main observations. First, the null vector condition is equivalent to the momentum-space holomorphicity of the cell periodic Bloch vectors $Q(k) [w_a \partial_k^a] \ket{u_k} = 0$, where $Q(k) = 1 - \ket{u_k}\bra{u_k}$ and $\partial_k^a = \partial / \partial k_a$~\cite{mera2021kahler}. Second, this holomorphicity condition stringently constrains the $\ket{\psi_k}$ to admit a universal form descending from a Landau level $\psi_k(r) \equiv e^{-\tilde\rho(r)} \phi_k (r) /N_k$~\cite{wang2021exact}, with $e^{-\tilde\rho(r)}$ a real positive and periodic function~\footnote{We can always choose a positive since only $|e^{-\tilde\rho}|$ appears in the overlaps and form factors of the Chern band~\cite{wang2021exact}, any additional phase can be absorbed in the boundary condition sewing function $s$.}, and $\ket{\phi_k} \in \mathcal{L}(g_{ab} = \Re [w_a w_b^*], B_0 = 2\pi/|a_1 \times a_2|)$ the element of a uniform LLL diagonalizing the magnetic translations by $a_{j=1,2}$ with eigenvalues $e^{i k_j}$. These magnetic translations commute due to the choice of $B_0$, which corresponds to having a single flux quantum threading each unit cell in the Landau problem. Finally, the scalar product remains the canonical one on the plane, such that $N_k^2 = \int {\rm d}^2 r \, |e^{-\tilde\rho} \phi_k|^2$.

Written in this form, the $q$-ideal band considered can be straightforwardly mapped onto the generalized LLL of Eq.~\ref{eq_mostgeneralLLL}. First, the linear transformation $Z = X+iY = w_a x^a$ provides the isothermal coordinates of the system, as it transforms the metric into $g_{ab} dx^a dx^b = dZ d\bar{Z}$, where, for the sake of simplicity, we have performed a global rescaling and rotation to fix $Z(a_1) = 1$. Then, the parameter $\tau$ of the torus is determined from the magnetic boundary conditions $\tau = Z(a_2)/Z(a_1)$~\cite{wang2021exact}. Finally, we observe that $B_0$ and $\tilde\rho$ are defined identically here and in Eq.~\ref{eq_mostgeneralLLL}, which completes the mapping of $q$-ideal Chern bands to LLLs with isotropic and uniform metric but spatially varying magnetic fields $\mathcal{L}(\delta_{ab}, B_0+\tilde{B})$, already hinted at in previous works~\cite{douglas2010bergman,konstantinou2017forgotten,dong2022dirac}.

\underline{Real-space condition.} The momentum-space condition for idealness has been generalized to a larger family of Chern bands~\cite{ledwith2022vortexability}, argued to go beyond the LLL mimicry of $q$-ideal bands. In more details, a band $\mathcal{C}$ is $r$-ideal if there exists a function $\mathfrak{z}:\mathbb{R}^2 \mapsto \mathbb{C}$, which, for periodic systems, must satisfy $\mathfrak{z}(r+ a_j) = \mathfrak{z}(r) + \mathfrak{z}(a_j)$, and if $\mathcal{C}$ is stable under multiplication by $\mathfrak{z}$. Due to its transformation under lattice translations, $\mathfrak{z}$ can be viewed as a non-linear real-space unit cell embedding, or equivalently, as a change of coordinates $F : r \to \mathfrak{r}$. Here, the term ``band'' underlies translation symmetry on a lattice with generating vectors denoted by $a_{1,2}$, and we may choose $\mathfrak{z}(a_1)=1$ and $\mathfrak{z}(a_2) = \tau$ as above.

Ref.~\cite{ledwith2022vortexability} observed a relation between $r$-ideal and $q$-ideal bands when using the $F$ non-linear embedding to modify the periodic part of the Bloch functions $\ket{u_k} = e^{-i (k, \mathfrak{r})} \ket{\psi_k}$, with $(k, \mathfrak{r}) = k_a M^{ab} \mathfrak{r}_b$ defined by the real invertible matrix $M^{ab}$ solving the two equations $M^{ab}  \mathfrak{r}_b (a_i) = a_i^a$. This matrix and its inverse $M_{ab}$ are introduced to avoid distorting the Brillouin zone. Then, the stability of $\mathcal{C}$ under $\mathfrak{z}$-multiplication becomes equivalent to the requirement of momentum-space holomorphicity for the $\ket{u_k}$. To see this, we recast the definition of $r$-idealness as $(1-P) \mathfrak{z} \ket{\psi_k} = 0$ with $P = \sum_k \ket{\psi_k}\bra{\psi_k}$, and use the relation $\mathfrak{r}_a \ket{\psi_k} = -i M_{ab} [\partial_k^b \ket{\psi_k} - e^{i (k, \mathfrak{r})} \partial_k^b \ket{u_k}]$ with the identity $(1-P) \partial_k^a \ket{\psi_k} = 0$  to get, after applying momentum conservation, $Q(k) [w_a \partial_k^a] \ket{u_k} = 0$ with $w_a = M_{1a} + i M_{2a}$. In particular, this implies that $r$-idealness is more general than $q$-idealness due to the non-linear embedding of the unit cell allowed in the definition of the periodic Bloch functions $\ket{u_k} = e^{-i (k, \mathfrak{r})} \ket{\psi_k}$. Quoting the result derived above for $q$-ideal bands, we deduce that the Bloch wavefunctions of $\mathcal{C}$ can be expressed as $\psi_k(r) \equiv e^{-\tilde\rho(\mathfrak{r})} \phi_k (\mathfrak{r}) / N_k$. The only difference with $q$-ideal bands being the normalization factor $N_k^2 = \int  {\rm d}^2 \mathfrak{r} \, |J_F| \cdot | e^{-\tilde\rho} \phi_k|^2$, which features the Jacobian $J_F$ of $F$.

This is sufficient to map $r$-ideal bands to the LLL of Eq.~\ref{eq_mostgeneralLLL}. Isothermal coordinates are obtained as in the $q$-ideal case through $Z = X+iY = w_a \mathfrak{r}^a$, where we similarly have imposed $Z(a_1)=1$ and defined $\tau = Z(a_2)$. To reproduce the normalization of $r$-ideal bands, we want the metric of the analog LLL to feature the same Jacobian factor $ds^2 = | J_F| (d X^2 + d Y^2)$, which is for instance accomplished using $g = |J_F (r)| \mathfrak{g}(r)$ with $\mathfrak{g}_{ab} = \Re[\partial_a \mathfrak{z}^* \partial_b \mathfrak{z}]$. Once the metric is known, $\tilde \rho$ and $\tilde B$ are obtained as in the $q$-ideal case. This concludes the mapping of $r$-ideal bands to the LLL $\mathcal{L}(g,B)$, where both $g$ and $B$ need to be periodic to capture the non-uniformity of the Berry curvature and the normalization of the more general $r$-ideal bands (see Tab.~\ref{tab_testideal} for summary).

\underline{Exhaustion of all periodic LLLs.} We have just proved that any ideal bands can be mapped to the most general form of a Landau level $\mathcal{L}(g, B)$ with periodic metric and magnetic field (Eq.~\ref{eq_mostgeneralLLL}). Conversely, 
any LLL $\mathcal{L}(g, B)$ is trivially an $r$-ideal band. Indeed, the $\mathcal{L}(g, B)$ are stable under multiplication by $Z= X+iY$ the complex isothermal coordinate, a property inherited from the flat and uniform LLLs spanned by the $\phi_k$ of Eq.~\ref{eq_flatuniformLLL}. In conclusion, ideal bands are in one-to-one correspondence with LLL in curved space with non-uniform magnetic fields. Consequently, there does not exist ideal bands beyond those already identified in the literature.

\paragraph*{Indiscriminate LLL mapping criterion --- } We now turn to the practical problem of probing LLL mapping in its most general form. In Tab.~\ref{tab_testideal}, we summarize the existing tests for checking whether a band is $q$- or $r$-ideal. The probe for the more restrictive $q$-idealness is computationally inexpensive. It only necessitates the calculation of the quantum geometric tensor over the full Brillouin zone, that is, the evaluation of the momentum derivatives of the $\ket{\psi_k}$. On the other hand, no efficient test of $r$-idealness exists at the moment. The only proposition towards finding one is to check the $q$-idealness condition with the quantum metric $\mathcal{Q}^F$ defined using $\ket{u_k^{F}} = e^{-i k \cdot F(r)} \ket{\psi_k}$ for all possible changes of coordinates $F$. This test should be understood as a formal description of the connection between $r$-ideal and $q$-ideal bands discussed above, more than an actual computational tool because of the unpractical minimization over all non-linear embeddings of the unit cell. 

\begin{table}
\centering
\begin{tabular}{c||c|c|c}
 Type & $g$ & $B$ & Test \\ \hline \hline
$q$-ideal & Uniform & Any & $\int_{BZ} \det \mathcal{Q}_k = 0$~\cite{wang2021exact} \\ \hline
$r$-ideal & Any & Any & ${\displaystyle \min_{F}} \int_{BZ} \det \mathcal{Q}_k^{F} = 0$~\cite{ledwith2022vortexability} 
\end{tabular}
\caption{Summary of the two types of ideal bands, the type of $\mathcal{L}(g,B)$ they map onto (see text), and the corresponding currently known test of idealness. For $r$-ideal bands, $\mathcal{Q}^{F}$ denotes the quantum geometric tensor computed using the $\ket{u_k^{F}}$ ensuing from the non-linear embedding of the unit cell $F$.}
\label{tab_testideal}
\end{table}

By explicit construction, we show that an operational test discriminating bands that map to the most general periodic LLL, and only involves derivatives of the Bloch wavefunctions, is possible. We first rationalize our criterion using some general properties of $\mathcal{L}(g,B)$, and then check that it indeed provides a criterion for LLL equivalence when applied to Chern bands. 

Starting from Eq.~\ref{eq_mostgeneralLLL}, and remembering that $\phi_k(X,Y)$ is a product of an analytical function in $Z=X+iY$ with a $k$-independent non-holomorphic form factor, we observe that
\begin{equation} \label{eq_difflogpsi}
\lambda_k : r \rightarrow  \psi_k (r) / \psi_{0} (r) ,
\end{equation}
is meromorphic in $Z$; the origin of the Brillouin zone $k=0$ can be arbitrarily chosen. The complex structure corresponding to this meromorphicity can be obtained as 
\begin{equation} \label{eq_complexstruct}
J_k = (d\lambda_k)^{-1} \cdot J \cdot d\lambda_k , \quad d\lambda_k = \begin{bmatrix} \partial_1 \Re \lambda_k &  \partial_2 \Re \lambda_k \\  \partial_1 \Im \lambda_k & \partial_2 \Im \lambda_k \end{bmatrix} ,
\end{equation}
of the complex plane's canonical $J_{ij} = \varepsilon_{ij}$. Note that the $J_k$ and invariant under the gauge transformation $\psi_k \to e^{i\theta_k} \psi_k$. Because $Z$ is momentum independent, so should the complex structure $J_k$ be. This statement forms the operational probe of LLL mapping that we propose 
\begin{equation} \label{eq_proposedcriterion}
\partial_k^a J_k = 0 . 
\end{equation}
A straightforward calculation shows that Eq.~\ref{eq_proposedcriterion} indeed holds for the example of $r$-ideal bands given in Ref.~\cite{ledwith2022vortexability}, a consequence of our ealier mapping of $r$-ideal bands to $\mathcal{L}(g,B)$ LLLs (see App.~\ref{app_checkcriterion}).

Let us now prove that bands fulfilling Eq.~\ref{eq_proposedcriterion} map onto a LLL $\mathcal{L}(g, B)$. In that case, each of the differentiable function $\lambda_k$ defined by Eq.~\ref{eq_difflogpsi} is holomorphic \emph{w.r.t.} the complex structure $J_k$ defined by Eq.~\ref{eq_complexstruct}, except where its poles lie. For generic physical systems, $\psi_{0}$ only features isolated zeros that do not produce essential singularities. The $\lambda_k$ are thus meromorphic \textit{w.r.t.} the complex structure $J_k$. In addition, if this complex structure does not depend on $k$, as described by Eq.~\ref{eq_proposedcriterion}, all $\lambda_k$ are meromorphic functions of the same complex variable $Z = X+iY$, which is only unique up to global conformal transformation. In particular, rotation and rescaling allows to set the image of $Z(a_1)=1$ and $Z(a_2)=\tau$ for a certain complex parameter $\tau$. The existence of a common complex variable $Z$ for all $\lambda_k$ allows for analytical progress since pseudo-periodic meromorphic functions on the torus are entirely determined by their boundary conditions and by the position of their zeros and poles.

To study these zeros and poles, we fix $k$ and introduce the auxiliary function $\mu_k = \partial_Z \log  \lambda_k = (\partial_Z \lambda_k) / \lambda_k$. It has simple poles with residue equal to $1$ and $-1$ for each of the ${\cal N}_z$ zeros and ${\cal N}_p$ poles of $\lambda_k$, respectively. The contour integral of $\mu_k$ over the unit cell is thus equal to the difference ${\cal N}_z-{\cal N}_p$. On the other hand, the periodicity of $\mu_k$ requires this integral to vanish. Therefore, $\lambda_k$ has as many poles as it has zeros ${\cal N}_z = {\cal N}_p = {\cal N}$. We respectively denote them as $Z_{0}^j$ and $Z_k^j$ with $j=1 \cdots {\cal N}$. These poles and zeros fully specify the functional form of the meromorphic functions
\begin{equation} \label{eq_meromorphicfunction}
\lambda_k (r) = C_k e^{i \alpha_k Z} \prod_{j=1}^{\cal N} \frac{\theta_1 (Z-Z_k^j; \tau)}{\theta_1 (Z-Z_{0}^j; \tau)} , 
\end{equation}
up to a constant $C_k$. To reproduce the boundary conditions $\lambda_k (r + a_j) = e^{i k_j} \lambda_k (r)$, the real parameter $\alpha_k$ and the center of mass of the zeros $S_k = (\sum_j Z_k^j)/\mathcal{N}$ must satisfy~\cite{NIST:DLMF}
\begin{equation} \label{eq_howzerosmove}
e^{i k_1} = e^{i\alpha_k} , \quad e^{i k_2} = e^{i \tau \alpha_k + 2i\pi\mathcal{N} (S_k-S_0)} .
\end{equation}

Assuming the Bloch eigenvectors to have no singularities for physical groundedness, the $Z_{k}$'s are a subset of the $\psi_k$'s zeros for each $k$. We dub them ``moving'' zeros as they change when $k$ is varied. Eq.~\ref{eq_howzerosmove} describes the motion of the center of mass these moving zeros as $k$ moves in the Brillouin zone. This motion is precisely that of the zero of LLL wavefunction on a torus (see App.~\ref{app_FormLLLelements}). We now consider different cases depending on the value of ${\cal N}$, and show that refined criteria can be obtained depending on the type of Landau level one wants to map onto.

A direct consequence of our criterion Eq.~\ref{eq_proposedcriterion} is that there must be at least one moving zero ${\cal N} \geq 1$. Indeed, ${\cal N}=0$ produces a constant $\alpha_k = 0$ that cannot satisfy the condition imposed by Eq.~\ref{eq_howzerosmove}. We then focus on the important case of a single moving zero ${\cal N}=1$. There, $Z_k = \alpha_k$ matches the position of the zero in a LLL on the torus. We realize that the ratio $\phi_k (X,Y)/\phi_0 (X,Y)$ possesses the same boundary condition, the same zeros and the same poles as $\lambda_k$. Both functions being meromorphic in $Z=X+iY$, they must be equal up to a constant. In other words, states of the bands can be written as 
\begin{equation} \label{eq_generalLLLmimicry}
\psi_k (r) \equiv \frac{e^{-\tilde\rho(r)}}{N_k} \phi_k (X,Y) , \quad e^{-\tilde\rho(r)} = \left| \frac{\psi_0(r)}{\phi_0(X,Y)} \right| ,
\end{equation}
where the phase of the ratio $\psi_0(r) / \phi_0(X,Y)$ does not change the overlaps nor the phase factors in the analog LLL, and can thus be absorbed into the boundary condition sewing function $s$. This brings us back to the case treated above for $r$-ideal bands, and hence completes the proof.

We have shown that our criterion Eq.~\ref{eq_proposedcriterion}, applied in a band with a single moving zero and where $\psi_{0} (r)$ only vanishes polynomially, implies LLL mapping in its most general form. We note that the two additional assumptions on the number of moving zeros and the absence of essential singularity are satisfied for $|C|=1$ bands obtained at all magic angles of the chiral model for twisted bilayer graphene~\cite{wang2021chiral}.

\paragraph*{More than one zero --- } Turning to a situation with more than one zero reveals a novel phenomenology of holomorphic bands beyond the most general LLL-mapping argument highlighted in this work. To see this, consider a single band satisfying Eq.~\ref{eq_proposedcriterion} and featuring $\mathcal{N}>1$ moving zeros. This band can only be identified as a strict subset of a $\mathcal{L} (g,B)$ with $N_\phi = \mathcal{N}$ flux per unit cell, the exact mapping between the two systems being proscribed by the larger $N_\phi$-fold degeneracy of LLL at each $k$-point (see App.~\ref{app_FormLLLelements}). While it is not clear whether this situation is compatible with a spectral gap isolating the band in energy, we were not able to rule it out. The properties of such a band would be determined by the specific cut it defines within the larger $\mathcal{L} (g,B)$ manifold, and cannot be solely inferred from the properties of the LLL. That is, holomorphic bands with $\mathcal{N}>1$ should go beyond the physics of a LLL.

From a different perspective, we see that such bands are incompatible with the $r$-idealness condition -- they are not stable under multiplication by $Z$. Should they be, the system would necessarily exhibit an $\mathcal{N}$-fold degeneracy at each point of the Brillouin zone, and precisely map onto an LLL featuring the same number of moving zeros (see App.~\ref{app_MoreThanOneZero}). This again highlights that $r$-idealness are equivalent to LLLs, even in the case of multiple bands.

\paragraph*{Greater Chern numbers ---} Let us finally comment on the case of a band with Chern number $|C|>1$. Early in the study of Chern bands, it was shown that such bands could be described as $|C|$ bands distinguished by a ``color'' index upon extending the unit cell $|C|$ times, where all the colored bands carry a Chern number ${\rm sign}(C)$ and are intertwined through real space translations by $a_{1/2}$, the original Bravais lattice vectors~\cite{barkeshli2012topological,wu2012gauge}. When these colored-bands took the form of ``flat ideal bands'' with Chern number $C_\sigma = 1$, \textit{i.e.} of Landau-levels with uniform magnetic field and metric, this construction provided flat ideal bands with larger Chern $|C|>1$ dubbed color-entangled~\cite{wu2014haldane}. More recently, $q$-ideal bands with $|C|>1$ were also written as colored-entangled bands where each of the colored band with unit Chern number was itself a $q$-ideal band~\cite{wang2022origin,dong2022exact}, \textit{i.e.} they mapped to Landau-levels with spatiallt varying magnetic field but uniform metric. 

One of the main idea of the present work is that the most generic ideal Chern band should ensue from the most generic Landau level, which possess both a non-uniform magnetic field and a non-uniform metric. Similarly, the most generic color-entangled ideal bands should be built from the Landau levels of Eq.~\ref{eq_mostgeneralLLL} including the non-linear embedding of of the unit cell provided by the isothermal coordinates $(X,Y)$. In App.~\ref{app_mostgenericcolorcentered}, we prove that applying the criterion Eq.~\ref{eq_proposedcriterion} to the colored bands of a generic bands with a Chern number $C>1$ yields precisely to this generalization. Thus, the framework presented here still applies provided we first disentangle the different `colors' of bands with greater Chern numbers.

\paragraph*{Conclusion --- } In this work, we have shown that all ideal Chern bands introduced in the literature map onto LLLs in curved space. Conversely, equivalence to the most general LLL does not lead to novel ideal bands beyond $r$-ideal bands. This proves that we have exhausted all possible criteria expressing Chern idealness as some flavor of LLL-mapping. We have also designed an operational criterion to identify which bands can be mapped onto the most general periodic LLLs. This criterion solely relies on the inexpensive evaluation of Bloch wavefunction's derivatives.

\paragraph*{Acknowlegments --- } We thank Jie Wang for useful comments. This work has benefited from discussions held at the 2023 Quantum Geometry Working Group meeting that took place at the Flatiron institute. The Flatiron Institute is a division of the Simons Foundation. N.R. acknowledges support from the QuantERA II Programme that has received funding from the European Union’s Horizon 2020 research and innovation programme under Grant Agreement No 101017733.

\bibliography{idealbands}

\appendix

\section{Global isothermal coordinates} \label{app_GlobalIsothermal}

The two-dimensional plane with the standard orientation and equipped with a Riemannian metric $g$ can be turned into a Riemann surface $\Sigma$, as the metric induces a complex structure.  By the uniformization theorem, $\Sigma$ is conformally equivalent to either the complex plane $\mathbb{C}$ or the unit disk $\mathbb{D}$. In this work, it is assumed that the metric $g$ is periodic, that is invariant under a discrete group $G$ of two-dimensional translations. Since they are metric-preserving, these translations are holomorphic with respect to the induced complex structure. The quotient space $\Sigma/G$ is a complex torus and thus isomorphic to $\mathbb{T}=\mathbb{C}/(\mathbb{Z}+\tau\mathbb{Z})$ for a certain complex parameter $\tau$  determined (up to modular transformations) by the metric $g$. Thus, $\Sigma$ is the universal covering space of the complex torus $\mathbb{T}$ and is therefore conformally equivalent to $\mathbb{C}$. There exists a conformal map $Z=X+iY:\mathbb{R}^2\to\mathbb{C}$, and $(X,Y)$ provides a \emph{global} isothermal coordinates, that is coordinates in which the metric has the form 
\begin{align}
g = e^{2 \sigma (X,Y)} (dX^2 + dY^2)
\end{align}
for some smooth function $\sigma$, which results in $g^{ab} = e^{-2 \sigma} \delta^{ab}$ and $\sqrt{g} = e^{2 \sigma}$ in the new coordinates. One convenient choice for the map $Z$ it to take the lift of the quotient map $\Sigma \to \mathbb{T}$. This ensures that $Z(r+a_1)=Z(r)+1$ and $Z(r+a_2)=Z(r)+\tau$, where $r\to r+a_1$ and $r\to r+a_2$ denote the generators of the translations on $\mathbb{R}^2$.

\section{Form of LLL elements Eq.~\ref{eq_mostgeneralLLL}} \label{app_FormLLLelements}

In this appendix we solve the Landau problem on the plane with a periodic setup as described in the main text. We shall denote as ``standard LLL'' those defined on a flat isotropic torus threaded by a uniform magnetic field.

\subsection{Reduction to the standard LLL}

Based on the previous appendix, we can work in coordinates $(X,Y)$ in which the metric is conformal 
\begin{align}
ds^2 = e^{2 \sigma (X,Y)} d Z \,d \bar{Z}
\end{align}
where $Z = X + i Y$ is a holomorphic coordinate, and both the metric and the magnetic field are invariant under $Z \to Z +1$ and $Z \to Z+ \tau$.  In isothermal coordinates the  Landau Hamiltonian of Eq.~\ref{eq_genericLandau} becomes $\mathcal{H}_{g, B} = \frac{1}{2 m} e^{-2 \sigma} \Pi \,\overline{\Pi}$ where $\Pi = \pi_X + i \pi_Y$, and LLL wavefunctions are those annhilated by $\bar{\Pi}$. This property is conformally invariant, in the sense that it is not sensitive to the conformal factor $e^{2\sigma}$. Thus, this factor does not modify the LLL wavefunctions, it only changes their inner product. 
Thus, it is sufficient to solve the LLL in the flat metric $ds^2 = dZ \,d \bar{Z}$.

Let $T_1$ and $T_2$ denote the magnetic translations operators corresponding to the elementary translation $Z \to Z+1$ and $Z \to Z+\tau$, respectively. A state $\ket{\psi_k}$ has quasi-momentum $k = (k_1,k_2)$ when  
\begin{align}
T_j \ket{\psi_k} = e^{i k_j} \ket{\psi_k} \,.
\end{align}
Thus LLL wavefunctions on the plane with a quasi-momentum $k= (k_1,k_2)$ coincide with LLL  wavefunctions on the torus $T = \mathbb{C}/({\mathbb{Z} + \tau \mathbb{Z}})$ in the presence of twists or magnetic fluxes $e^{i k_1}$ and $e^{i k_2}$.

In order to be more explicit we now make a gauge choice. We first decompose the total magnetic field as $B = B_0 + \tilde{B}$ into a uniform part $B_0$ and a periodic part $\tilde{B}$ that carries no flux
\begin{align}
B_0 = \frac{2\pi N_\phi}{\textrm{Im} \tau}, \qquad  \int_{\textrm{unit cell}} \tilde{B}  \, dX dY =0 \,.
\end{align}
where the integer $N_\phi$ is the number of flux quanta per unit cell. Correspondingly we introduce the gauge $A_Z = - i \partial_Z \rho$, $A_{\bar{Z}} = i \partial_{\bar{Z}} \rho$ where we chose for the K\"ahler potential $\rho = \rho_0 + \tilde{\rho} (Z,\bar{Z})$ with $\rho_0 = B_0 Z \bar{Z}/4$, and where $\tilde{\rho}$ is periodic and satisfies $\Delta \tilde{\rho}  = \tilde{B}$ in the flat metric $ds^2 = dZ \,d \bar{Z}$, that is $\Delta \tilde{\rho}= \tilde{B}/\sqrt{g}$ in the original metric. Since $\tilde{\rho}$ is periodic, it is insensitive to magnetic translations, and simply factors out. As a result, we get~\cite{douglas2010bergman,konstantinou2017forgotten,dong2022dirac}
\begin{equation}
\psi_{k,n}(Z,\bar{Z}) = \phi_{k,n} (Z,\bar{Z}) e^{-\tilde{\rho} (Z,\bar{Z})} / N_k 
\end{equation}
where $\phi_{k,n}(Z,\bar{Z})$ denotes the standard LLL wavefunctions on a flat torus, in the presence of a uniform magnetic field $B_0$, and with magnetic fluxes $e^{ik_1}$ and $e^{ik_2}$. Therefore there are $n=1,\cdots ,N_\phi$ linearly independent LLL states for each value of $k$.

\subsection{Wavefunctions of the standard LLL}

We now recall the form of the standard LLL wavefunctions. In the chosen gauge, specified by $\rho_0$, generic LLL states are of the form $\phi(Z,\bar{Z}) = f(Z) e^{-\rho_0(Z,\bar{Z})+\rho_0(Z,Z)}$, with $f$ holomorphic in the variable $Z$ and where the holomorphic exponential term $e^{\rho_0(Z,Z)}$ is conventional. Imposing quasi-periodic boundary conditions compatible with quasi-momentum $k=(k_1,k_2)$ further demands 
\begin{equation} \label{eq_magnetic_translations} \begin{split}
\phi_k (Z + R) & =   e^{i \frac{B_0}{2}  \textrm{Im}(\bar{R}Z) + i ( k_1 n_1 + k_2 n_2 ) } e^{i \pi N_\phi n_1 n_2 }  \phi_k (Z) , \\
f_k(Z+R) & = e^{-i \pi n_2 N_\phi (2Z + n_2 \tau) + i ( k_1 n_1 + k_2 n_2 ) }  f_k(Z) ,
\end{split} \end{equation}
for any $R = n_1  + n_2 \tau$ in the lattice. It is a standard result that such quasi-periodic holomorphic functions have exactly $N_\phi$ zeros per unit cell (counted with multiplicites), denoted here as $Z_k^j$ with $j=1,\cdots,N_\phi$. They can furthermore be factorized as 
\begin{equation} \label{appeq_factorLLL}
f_k (Z) = C_k e^{i \alpha_k Z} \prod_{j=1}^{N_\phi} \theta_1(Z-Z_k^j ; \tau) ,
\end{equation}
with $\theta_1$ the first Jacobi theta function
\begin{equation} \label{appeq_theta_1}
\theta_1(Z ; \tau)  = \sum_{n \in \, \mathbb{Z}+1/2} e^{\pi i \tau n^2} e^{2\pi i n (Z-1/2)}\,,
\end{equation}
where the real constant $\alpha_k$ is the unique solution of~\cite[chap.~20]{NIST:DLMF}
\begin{equation} \begin{split}
e^{i k_1} & =  (-1)^{N_{\phi}}e^{i \alpha_k}, \\
e^{i k_2} & = (-1)^{N_{\phi}}e^{ i \alpha_k \tau} e^{2\pi i \sum_j Z_k^j} .
\end{split} \end{equation}
and the zeros can be anywhere provided they satisfy the sum rule 
\begin{equation}
\sum_{j=1}^{N_\phi} Z_k^j = \frac{k_2 - \tau k_1}{2\pi} + \frac{N_{\phi}}{2}(1 - \tau)
\end{equation}
modulo $\mathbb{Z} + \tau \mathbb{Z}$. The Rieman-Roch theorem ensures that we can build $N_\phi$ linearly independent solutions to these equations, and that their linear combinations can generate any function of the form given in Eq.~\ref{appeq_factorLLL}.

We now briefly come back to $N_\phi = 1$, which is the main focus of the main text. The previous discussion shows that there is a single wavefunction, whose unique zero must be located at 
\begin{equation}
Z_k = \frac{(k_2 +\pi) - \tau (k_1+\pi)}{2\pi} , 
\end{equation}
in the unit cell $\{ r_1+r_2 \tau | (r_1,r_2)\in [0,1)^2 \}$, leading to $\alpha_k = k_1+\pi$. This fully specifies the functional form of LLL states (up to a normalization constant) to the one given in the main text, and repeated here for clarity
\begin{equation}
\phi_k (X,Y) = e^{\frac{B_0}{4} Z (Z - \bar{Z}) + i (k_1 +\pi) Z } \theta_1 (Z-Z_k; \tau) .
\end{equation}

\section{Criterion applied to known ideal bands} \label{app_checkcriterion}

In this appendix, we show that the criterion Eq.~\ref{eq_proposedcriterion} is satisfied for the isolated ideal bands studied in Refs.~\cite{ledwith2022vortexability,wang2021exact,crepel2023chiral}, which include generic $q$-ideal bands, periodically strained graphene, twisted bilayer graphene and transition metal dichalcogenides in the chiral limit. In these systems, the Bloch eigenfunctions are respectively normalized square integrable functions, two-components spinors with a single non-zero entry, or four-component spinors with two non-zero entries related to one another by a symmetry of the model. As a result, the Bloch states in these systems can all be described by single scalar functions, which for spinors correspond to the topmost non-zero entry. Furthermore, in all these cases, the ratios $\lambda_k$ defined by Eq.~\ref{eq_difflogpsi} are found to only depend on a single variable $Z=X+iY$ independent of the quasi-momentum $k$~\cite{ledwith2022vortexability,wang2021exact}. In the case of strained graphene, the variable $Z$ is not necessarily equal to $z=x+iy$, the real-space holomorphic coordinate in the lab frame. 

Using the short-hand notations $D_k = \det (d\lambda_k)$, $R_k = \Re (\lambda_k)$ and $I_k = \Im (\lambda_k)$, we find that Eq.~\ref{eq_complexstruct} can be recast as 
\begin{equation} \label{appeq_CheckCriterion} \begin{split}
(J_k)_{ij} & = \varepsilon_{ip} \frac{\partial_p I_k \partial_j I_k+\partial_p R_k \partial_j R_k}{D_k} \\
& = i \varepsilon_{ip} \frac{\partial_p \lambda_k \partial_j \lambda_k^* + \partial_p \lambda_k^* \partial_j \lambda_k}{\partial_1\lambda_k \partial_2 \lambda_k^* - \partial_1\lambda_k^* \partial_2 \lambda_k} \\
& = i \varepsilon_{ip} \frac{|\lambda_k'|^2}{|\lambda_k'|^2} \frac{\partial_p Z \partial_j Z^* + \partial_p Z^* \partial_j Z}{\partial_1 Z \partial_2 Z^* - \partial_1 Z^* \partial_2 Z} \\
& = \varepsilon_{ip} \frac{\Re [\partial_p Z \partial_j Z^*]}{\Im [\partial_1 Z \partial_2 Z^*]}, 
\end{split} \end{equation}
where we have applied the chain rule to get the third line and denoted the derivative of $\lambda_k$ with respect to $Z$ as $\lambda_k'$. The latter only depends on $Z$ but not on $k$, and our criterion Eq.~\ref{eq_proposedcriterion} is satisfied.

The first line in Eq.~\ref{appeq_CheckCriterion} also allows to check that the $J_k$ are invariant under a gauge transformation $\lambda_k \to e^{i\theta_k } \lambda_k$ which changes 
\begin{equation}
\begin{bmatrix} R_k \\ I_k \end{bmatrix} \to \begin{bmatrix}  \cos \theta_k & -\sin \theta_k \\ \sin\theta_k & \cos\theta_k \end{bmatrix} \begin{bmatrix} R_k \\ I_k \end{bmatrix} , 
\end{equation}
but sends
\begin{equation}
\partial_p I_k \partial_j I_k+\partial_p R_k \partial_j R_k \to \partial_p I_k \partial_j I_k+\partial_p R_k \partial_j R_k ,
\end{equation}
for any value of $p$ and $j$, and therefore leaves $J_k$ invariant. 

Let us finally comment that Ref.~\cite{ledwith2022vortexability} also considers cases where $n$ independent bands, with Chern number equal to one, are grouped together to obtain a system carrying a total Chern number $C=n$. This construction should be contrasted with the case of a \textit{single} ideal Chern band -- the main focus of our paper -- with Chern number $C>1$ considered in App.~\ref{app_mostgenericcolorcentered}. We note, however, that the $n$ bands in the systems considered in Ref.~\cite{ledwith2022vortexability} are obtained from a single square integrable non-spinor wavefunction and a specific recursive scheme allowing to infer the wavefunctions of the $n$ bands from this single wavefunction without spoiling the complex structure (only anti-holomorphic derivative appear). The square integrable non-spinor wavefunction at the origin of this construction only depends on a complex variable $Z$ and therefore satisfies our criterion, as can be checked along the lines of Eq.~\ref{appeq_CheckCriterion}.

\section{More than one zero} \label{app_MoreThanOneZero}

In this appendix, we show that \emph{a single band} satisfying Eq.~\ref{eq_proposedcriterion} and featuring $\mathcal{N}>1$ moving zeros is not compatible with the $r$-idealness, or ``vortexability'', condition. Enforcing this condition requires the addition of $\mathcal{N}-1$ other bands.  Together, these $\mathcal{N}$ bands map onto a LLL $\mathcal{L}(g,B)$ with the same number of moving zeros, \textit{i.e.} threaded by $N_\phi = \mathcal{N}$ flux quanta per unit cell -- see App.~\ref{app_FormLLLelements}. 

Let us assume that a holomorphic band with $\mathcal{N}>1$ moving zeros is also stable under mutliplication by $Z$. This implies that for any analytic function $f(Z)$ and Bloch eigenvector $\psi_k (r)$ in the band, $f(Z) \psi_k (r)$ is also in the band. For the ratios of Eq.~\ref{eq_meromorphicfunction}, this ensures that $\lambda_k$ multiplied by a meromorphic function that does not change the number of zeros still describes the ratio between elements of the band. Choosing any set $\{ W_k^j \}$ such that $\sum_j W_k^j = S_k$, we can for instance obtain 
\begin{equation} \begin{split}
\hat\lambda_k(r) &= \lambda_k(r) \prod_{j=1}^{\mathcal{N}} \frac{\theta_1(Z-W_k^j ;\tau)}{\theta_1(Z-Z_k^j;\tau)} \\ &= C_k \frac{e^{i \alpha_k Z}}{N_k} \prod_{j=1}^{\mathcal{N}} \frac{\theta_1(Z-W_k^j;\tau)}{\theta_1(Z-Z_0^j;\tau)} , 
\end{split} \end{equation}
which describes the ratio $\hat\psi_k/\hat\psi_0$. There, $\hat\psi_0$ has moving zeros at the poles of $\hat\lambda_k$, located at $Z_0^j$ and is therefore proportional to $\psi_0$. Similarly, $\hat\psi_k$ vanishes at $W_k^j$, has quasi-momentum $k$ because its zeros' center of mass is equal to that of $\psi_k$, and must belong to the band. Since there is only one state with quasi-momentum $k$ in the band considered, it must be proportional to $\hat\psi_k = a \psi_k$. However, since $\mathcal{N}>1$, we can choose the new zeros such that one of them $W_k^{j_0}$ (at least) does not appear in set of original zeros $\{Z_k^j\}$. This yields the contradiction $\hat\psi_k (W_k^{j_0}) = 0 \neq a \psi_k(W_k^{j_0})$.

Therefore, $r$-idealness can only be fulfilled if $\mathcal{N}>1$ when our criterion is extended to several bands, \textit{i.e.} if we replace Eq.~\ref{eq_difflogpsi} by $\lambda_{k,n} = \psi_{k,n}/\psi_{0,0}$, with $1 \leq n \leq N_b$ a band index, and require $\partial_k^b J_{k,n} = 0$ such that the $\lambda_{k,n}$ remain of the form given in Eq.~\ref{eq_meromorphicfunction} with zeros located at $Z_{k,n}^j$ ($j=0,\cdots, \mathcal{N}-1$). Requiring the set of $N_b$ bands to be stable under $Z$ multiplication, and adapting the previous argument to the present situation, we realize that any function of the form of $\hat\lambda_k$ above with $\sum_j W_k^j = S_{k,0}$ must be a linear combination of the $\{\lambda_{k,n}\}_{n=1,\cdots ,N_b}$. On the other hand, an instance of the Riemann-Roch theorem assures that the set of $\hat\lambda_k$ satisfying the boundary conditions deduced from $\sum_j W_{k,n}^j = S_{k,0}$ (see Eq.~\ref{eq_howzerosmove} in main text) is fully generated by $\mathcal{N}$ linearly independent functions~\cite{crepel2019matrix}. This proves that $N_b \geq \mathcal{N}$. Furthermore, the holomorphicity condition ensures that the original $\lambda_{k,n}$ have the form given above for $\hat\lambda_k$, and thus $N_b = \mathcal{N}$ for the $N_b$ bands to be linearly independent. Finally, transposing the demonstration leading to Eq.~\ref{eq_generalLLLmimicry} in the main text to the $N_b = \mathcal{N}$ multiband case, we find $\psi_{k,n} \equiv \mathcal{B}(r) \phi_{k,n} (X,Y) / N_{k,n}$ with $\phi_{k,n}$ denoting elements of a LLL with a unit cell threaded by $N_\phi = \mathcal{N}$ flux quanta (see App.~\ref{app_FormLLLelements}), which completes the proof.

\section{$|C|>1$ as $|C|$ copies of $C=1$} \label{app_mostgenericcolorcentered}

\subsection{Definition and goal}

In this appendix, we consider the case of a band with Chern number $|C|>1$. Early in the study of Chern bands, it was shown that such bands could be described as $|C|$ bands with Chern number ${\rm sign}(C)$ upon extending the unit cell $|C|$ times and imposing specific boundary condition intertwining the $|C|$ bands together in real space~\cite{barkeshli2012topological}. The band index appearing due to this folding of the Brillouin zone is often named `color'~\cite{wu2012gauge}. In the cases where the $|C|$ bands on the extended unit cell can be mapped onto a Landau level with uniform metric and magnetic field, the original band with Chern number $C$ was dubbed color-entangled Landau level~\cite{wu2014haldane}. This was generalized in Refs.~\cite{wang2022origin,dong2022exact} to the case where the $|C|$ bands on the extended unit cell are $q$-ideal. The resulting color-entangled $q$-ideal bands $\{ \chi_k (r) \}$ with Chern number $C>1$ (we shall assume $C>0$ from now on) were shown to assume the form
\begin{equation} \label{appeq_colorentangledidealbands}
\chi_k (r) = \frac{1}{N_k} \sum_{m=0}^{C-1} e^{-i m k_1} \psi_k^{q\text{-id}} (r + m a_1 ) , 
\end{equation}
with $N_k$ a normalization coefficient, $(a_1,a_2)$ the original unit cell, and $\psi_k^{q\text{-id}}$ the wavefunctions of a $q$-ideal band with Chern number one defined on the enlarged unit cell $(|C|a_1,a_2)$. We recall that $\psi_k^{q\text{-id}}$ maps onto a lowest Landau level with non-uniform magnetic field by uniform metric.

One of the main idea of the present work is that the most generic ideal Chern band ensues from the most generic Landau level, which possesses both a non-uniform magnetic field and a non-uniform metric. In other words, the most generic color-entangled ideal bands should be built from the Landau levels of Eq.~\ref{eq_mostgeneralLLL} including the non-linear embedding of of the unit cell provided by the isothermal coordinates $(X,Y)$, \textit{i.e.} including a non-uniform metric. Our goal here is to adapt the mapping of Refs.~\cite{barkeshli2012topological,wu2012gauge} to this case and show that our criterion Eq.~\ref{eq_proposedcriterion} provides the desired generalization of Eq.~\ref{appeq_colorentangledidealbands}.

\subsection{Decomposition into colors, and back}

Let us consider a band with Chern number $C>1$ spanned by the Bloch functions $\{ \chi_k (r) \}$. It satisfies
\begin{equation}
\ket{\chi_{k+2\pi g_1}} = e^{2 i \pi \theta_k^{(1)}} \ket{\chi_{k}} , \quad \ket{\chi_{k+ 2\pi g_2}} =  e^{2 i \pi \theta_k^{(2)}} \ket{\chi_{k}} ,
\end{equation}
with $g_i \cdot a_j = \delta_{ij}$ and $k$ in the Brillouin zone $$BZ=\{k=k_1 g_1 + k_2 g_2 \, | \, (k_1, k_2) \in [0,2\pi)^2 \} , $$ where the phases $\theta_k^{(j)}$ are constrained by the Chern number
\begin{equation}
\theta_{k+ 2\pi g_2}^{(1)} - \theta_k^{(1)} + \theta_{k}^{(2)} - \theta_{k+ 2\pi g_1}^{(2)} = C ,
\end{equation}
but is otherwise arbitrary. For concreteness, we pick a gauge in which the $\theta_k^{(1)} = \theta(k_2)$ only depends on $k_2$ and $\theta_k^{(2)}=0$ vanishes. This choice is always possible; and encompasses the gauge choice used in Refs.~\cite{wang2022origin,dong2022exact} for the same folding of the Brillouin zone. The Chern number number now corresponds to the winding of the $\theta$ function $\theta(k_2 + 2\pi) = \theta(k_2) + C$. 

We define the wavefunctions of the colored bands as 
\begin{equation} \label{appeq_differentcolors}
\ket{\phi_{\sigma,k}} = \frac{1}{\sqrt{C}} \sum_{m=0}^{C-1} e^{2 i \pi m [\sigma - \theta (k_2)] / C} \ket{\chi_{k + 2 \pi m g_1/C}} ,
\end{equation}
with $\sigma = 0, \cdots , C-1$ and where $k$ now belongs to the folded Brillouin zone $$BZ_f = \{k=k_1 g_1 + k_2 g_2 | k_1 \in [0,2\pi/C), k_2\in [0,2\pi)\}$$ corresponding to an enlarged real-space unit cell spanned by $(C a_1,a_2)$. We find that the colored bands obey
\begin{equation} \begin{split}
\ket{\phi_{\sigma,k+2 \pi g_1/C}} & = e^{2i\pi [\theta (k_2)-\sigma]/C} 
\ket{\phi_{\sigma,k}} ,  \\  
\ket{\phi_{\sigma,k+2 \pi g_2}} & = \ket{\phi_{\sigma,k}}, 
\end{split} \end{equation}
which grants the colored bands a unit Chern number $C_\sigma = 1$. Finally, using the periodicity of the original band $\chi_k(r+a_j) = e^{i k_j} \chi_k(r)$, we derive the additional relations
\begin{equation} \begin{split}
\phi_{\sigma,k} (r+a_1) & = e^{i k_1} \phi_{\sigma+1,k} (r) , \\ \phi_{\sigma,k} (r+a_2) & = e^{i k_2} \phi_{\sigma,k} (r) , 
\end{split} \end{equation}
from which the enlarged periodicity $(|C|a_1,a_2)$ becomes evident (note also the similarity with Ref.~\cite{wu2014haldane}).

We can Fourier transform back to invert Eq.~\ref{appeq_differentcolors} and obtain the original Bloch functions in the form
\begin{equation}
\chi_{k + 2\pi m g_1/C} (r) = \frac{1}{\sqrt{C}} \sum_{\sigma=0}^{C-1} e^{-2i\pi m [\sigma - \theta(k_2)]/C} \phi_{\sigma, k} (r) .
\end{equation}
Using the boundary conditions derived above for the colored bands, we get
\begin{align} \label{appeq_originalWFintermsofcolors}
& \chi_{K} (r) = \frac{1}{\sqrt{C}} \sum_{\sigma=0}^{C-1} e^{2i\pi m [\theta(K_2) - \sigma]/C} \phi_{\sigma,K-2 \pi mg_1/C} (r) \notag \\ 
& = \frac{1}{\sqrt{C}} \sum_{\sigma=0}^{C-1} \phi_{\sigma,K} (r) = \frac{1}{\sqrt{C}} \sum_{\sigma=0}^{C-1} e^{- i \sigma K_1} \phi_{0,K} (r + \sigma a_1)  , 
\end{align}
with $K \in BZ$ and $K-mg_1/C$ the quotient of $K$ by $g_1/C$ defining $m$.

\subsection{Criterion}

Comparing Eq.~\ref{appeq_originalWFintermsofcolors} to Eq.~\ref{appeq_colorentangledidealbands} offers a transparent interpretation of the result of Refs.~\cite{wang2022origin,dong2022exact}: the colored wavefunction $\phi_{0,k}$ is a $q$-ideal band if the original band is. We can obtain the desired generalization of colored-entangled ideal bands that accounts for a non-uniform metric by simply applying our criterion Eq.~\ref{eq_proposedcriterion} to $\phi_{0,k}$. More explicitly, if $\phi_{0,k}$ verifies Eq.~\ref{eq_proposedcriterion}, our derivation in the main text and Eq.~\ref{appeq_originalWFintermsofcolors} shows that the original band of Chern number $C>1$ takes the form 
\begin{equation}
\chi_k (r) = \frac{1}{\sqrt{C}} \sum_{m=0}^{C-1} e^{-i m k_1} \psi_k (r + m a_1 ) ,
\end{equation}
where $\psi_k$ assumes the form given in Eq.~\ref{eq_generalLLLmimicry} on the lattice with extended periodicity $(|C|a_1,a_2)$. Note that the boundary conditions intertwining the different colors ensure that all $\phi_{\sigma \neq0,k}$ satisfy our criterion when $\phi_{0,k}$ does. 

\end{document}